\documentclass[fleqn,usenatbib]{mnras}

\usepackage{mathptmx}
\usepackage[T1]{fontenc}
\usepackage{ae,aecompl}
\usepackage{rotating}
\usepackage{graphicx}	% Including figure files
\usepackage{amsmath}	% Advanced maths commands
\usepackage{amssymb}	% Extra maths symbols
\usepackage{geometry}
\usepackage[figure,figure*]{hypcap}
\usepackage{epstopdf}

\newcommand{\raa}{Research in Astronomy and Astrophysics}
\title[17eaw,mnras]{Probing the Final-stage Progenitor Evolution for Type IIP Supernova 2017eaw in NGC 6946}

\author[Rui LM et al.]{
Liming Rui,$^{1}$
Xiaofeng Wang,$^{1}$%\thanks{E-mail:xf_wang@mails.tsinghua.edu.cn}
Jun Mo,$^{1}$
Danfeng Xiang,$^{1}$
Jujia Zhang,$^{2,3}$
\newauthor
Justyn R. Maund,$^{4}$
Avishy Gal-Yam,$^{5}$
Lifan Wang,$^{6}$
Tianmeng Zhang$^{7}$
\\
$^{1}$Physics Department and Tsinghua Center for Astrophysics, Tsinghua University, Beijing, 100084, China\\
$^{2}$Yunnan Observatories, Chinese Academy of Sciences, Kunming 650011, China\\
$^{3}$Key Laboratory for the Structure and Evolution of Celestial Objects, Chinese Academy of Sciences, Kunming 650216, China\\
$^{4}$Department of Physics and Astronomy, University of Sheffield, Hicks Building, Hounsfield Road, Sheffield S3 7RH, UK\\
$^{5}$Department of Particle Physics and Astrophysics, Weizmann Institute of Sciencem Rehovot 76100, Israel\\
$^{6}$Physics and Astronomy Department, Texas A\&M University, College Station, TX 77843, USA\\
$^{7}$Key Laboratory of Optical Astronomy, National Astronomical Observatories, Chinese Academy of Sciences, Beijing 100012,China
}

\date{Accepted XXX. Received YYY; in original form ZZZ}

\pubyear{XXX}

\begin{document}
\label{firstpage}
\pagerange{\pageref{firstpage}--\pageref{lastpage}}
\maketitle

\begin{abstract}
%This is a simple template for authors to write new MNRAS papers.
%The abstract should briefly describe the aims, methods, and main results of the paper.
%It should be a single paragraph not more than 250 words (200 words for Letters).
%No references should appear in the abstract.
We presented a detailed analysis of progenitor properties of type IIP supernova 2017eaw in NGC 6946, based on the pre-explosion images and early-time observations obtained immediately after the explosion. An unusually red star, with M$_{F814W}$ = $-$6.9 mag and m$_{F606W}$$-$ m$_{F814W}=$2.9$\pm$0.2 mag, can be identified at the SN position in the pre-discovery \textit{Hubble Space Telescope}(HST) images taken in 2016. The observed spectral energy distribution of this star, covering the wavelength of 0.6-2.0$\ \mathrm{\mu m}$, matches that of an M4-type red supergiant (RSG) with a temperature of about 3550 K. These results suggest that SN 2017eaw has a RSG progenitor with an initial mass of 12$\pm$2 M$_\odot$. The absolute F814W-band magnitude of this progenitor star is found to evolve from $-$7.2 mag in 2004 to $-$6.9 mag in 2016. Such a dimming effect is, however, unpredicted for a RSG in its neon/oxygen burning phase when its luminosity should modestly increase. The spectrum of SN 2017eaw taken a few hours after discovery clearly shows a narrow H$_\alpha$ emission feature blueshifted by $\sim$160 km s$^{-1}$. This narrow component disappeared in the spectrum taken two days later, suggesting the presence of a circumstellar material (CSM) shell (i.e., at a distance of $<$2.1-4.3$\times$10$^{14}$ cm). Combining the inferred distance with the expansion velocity of the CSM, we suggest that the progenitor of SN 2017eaw should have experienced violent mass loss at about 1-2 years prior to explosion, perhaps invoked by pulsational envelop ejection. This mechanism may help explain its luminosity decline in 2016 as well as the lack of detections of RSGs with initial mass in the range of 17 M$_\odot<$ M $<$ 25 M$_\odot$ as progenitors of SNe IIP.
\end{abstract}


%%%%%%%%%%%%%%%%%%%%%%%%%%%%%%%%%%%%%%%%%%%%%%%%%%
%\onecolumn
%%%%%%%%%%%%%%%%% BODY OF PAPER %%%%%%%%%%%%%%%%%%

\section{Introduction}
Massive stars with initial masses ranging from 8 to 25 M$_\odot$ usually end their lives as type II core-collapse supernovae \citep{2003ApJ...591..288H}. Among them, type IIP supernovae (SNe IIP) are the dominant subclass, which are characterized by prominent Balmer lines in the spectra and a plateau phase lasting for about 100 days in the light curves \citep{1979AA....72..287B, 1997ARAA..35..309F,2017suex.book.....B}. These hydrogen-rich SNe are identified to arise from core-collapse of red supergiants (RSGs) with initial mass lying between 8 $-$ 16.5 M$_{\odot}$ \citep[and reference therein]{2007ApJ...661.1013L,2008MNRAS.387.1344M,2009ARAA..47...63S}. However, the observation of RSGs in Milky Way, Large Magellanic Cloud(LMC) and Small Magellanic Cloud(SMC) indicate a mass range of RSGs from 9 $-$ 25M$_\odot$ \citep{2005ApJ...628..973L,2006ApJ...645.1102L,2007ApJ...667..202L}.  The lack of RSGs with mass in the range of 17 $-$ 25 M$_\odot$ as progenitors of SNe IIP challenges current theory of stellar evolution. One possible explanation is that some massive RSGs experience prominent mass loss during its final stage evolution towards the explosion, due to the stellar wind, binary interaction, or pulsational eruption. The accumulated circumstellar materials (CSM) may obscure the RSGs and lead to the underestimation of the progenitor luminosity and its initial mass. The condensed dust shell will leave an imprint on the SN spectra via photoionization or interaction with the SN ejecta \citep[e.g. ][]{2017NatPh..13..510Y,2018ApJ...861...63H}.

Signatures for the presence of nearby CSMs have been reported for SNe IIP/IIL/IIb. Early spectra of type IIb SN 2013cu \citep{2014Natur.509..471G} and type IIP SN 2013fs \citep{2017NatPh..13..510Y} reveal flash-ionized emission lines from dense circumstellar wind. The narrow emission lines are also detected in the early-time spectra taken in few hours or days after explosion, i.e., SN 2016bkv \citep{2018ApJ...861...63H}, SN 2006bp \citep{2007ApJ...666.1093Q} and some young type IIP/IIL SNe reported by \cite{2016ApJ...818....3K}. The narrow interacting features could be generated by recombination of the CSMs ionized by ultraviolet (UV) radiation emitted during the shock breakout or the shock-cooling phase from the supernovae. However, the mechanism driving the mass loss of RSGs, i.e., whether the mass loss is stable or unstable, is not fully understood. Dense CSM surrounding a typical type IIP supernova is found to have a expansion velocity that is apparently higher than typical wind velocity of RSGs  \citep{2017NatPh..13..510Y} (i.e., 130 km s$^{-1}$ versus 10 km s$^{-1}$), suggestive of eruptive mass loss during the late phase evolution of RSGs. Moreover,  \cite{2010ApJ...717L..62Y} found that strong pulsation induced by partial ionization of hydrogen in the envelope is expected to enhance the mass loss during the RSG phase.

SN 2017eaw provides another rare opportunity to study the late-time evolution of the red supergiant with relatively low initial mass.
This SN was discovered by Patrick Wiggins on 2017 May 14.238 (UT dates) in NGC 6946 at a distance of 5.5 Mpc, with an unfiltered CCD magnitude of 12.8 mag  \citep{2017CBET.4390....1W}, and it was immediately classified as a young type IIP supernova \citep{2017ATel10376....1X}. Note that NGC 6946 is a well-known nearby Scd galaxy that has recorded nine SNe before SN 2017eaw, which makes it one of the most prolific SN factories known in the local universe. Follow-up observations of SN 2017eaw were thus initiated less than 0.6 d after the discovery, including multi-color ultraviolet/optical photometry and rapid high-cadence spectroscopy. The UBVRI-band photometric evolution covering the first 200 days has been presented by \cite{2018arXiv180100340T}, which shows a long plateau characterized by a normal type IIP SN. \cite{2018MNRAS.481.2536K} studied the progenitor properties of SN 2017eaw, suggesting that it has a red supergiant progenitor with a 13 M$_{\odot}$. Based on the historical Spitzer data, their studies also suggested that the progenitor star might have experienced dust-shell formation a few years before the explosion.

In this paper, we also attempt to constrain the progenitor properties of type IIP supernova 2017eaw using photometric and spectroscopic observations obtained immediately after explosion and the archival images from HST and Spitzer. The description of early-time observation and brief data reduction process are presented in Section \ref{sec:snobs}. In Section \ref{sec:proobs}, we describe the identification of the progenitor candidate from HST and Spitzer in detail. We discuss the progenitor properties and explore its possible final-stage evolution before explosion in Section \ref{sec:alys}. Our conclusions are given in Section \ref{sec:sum}.
%The observations of the normal type IIP supernova 2017eaw and analysis of its preexplosion progenitor evolution indicate that a %relatively low-mass red supergiant may also experience significant mass ejection within a few years before its death.

\section{Early-time Followup Observations}
\label{sec:snobs}
The UBVRI photometry observations of SN 2017eaw were triggered immediately after the SN discovery using the Tsinghua-NAOC 80-cm telescope(TNT) \citep{2012RAA....12.1585H} at Xinglong Observatory of NAOC (Located in Hebei, China), with the first observations obtained at t$\sim$0.6 days from discovery. The SN was observed by the Swift UVOT \citep{2004ApJ...611.1005G} even earlier in three ultraviolet (UV) (\textit{uvw2, uvm2, uvw1}) and three optical filters (U, B, V), starting at t$\sim$0.4 days after discovery. Based on the explosion time deduced below, these two phases correspond to 1.6 days and 1.4 days after explosion, respectively.

For data obtained by TNT, the data were first processed with IRAF\footnote{IRAF is distributed by the National Optical Astronomy Observatories, which are operated by the Association of Universities fo	Research in Astronomy, Inc., under cooperative agreement with the National Science Foundation (NSF).}, this includes bias subtraction and flat field correction. The instrumental magnitudes were measured with point-spread function (PSF) photometry and photometric calibration was determined with 10 stars in the field of SN 2017eaw observed on photometric night. We adopted aperture photometry to reduce all Swift images using HEASoft(the High Energy Astrophysics Software)\footnote{\url{http://www.swift.ac.uk/analysis/software.php}} and detailed data descriptions are addressed in Rui et al. (2019; in prep).

\begin{figure*}
	\centering
	\includegraphics[width=\linewidth]{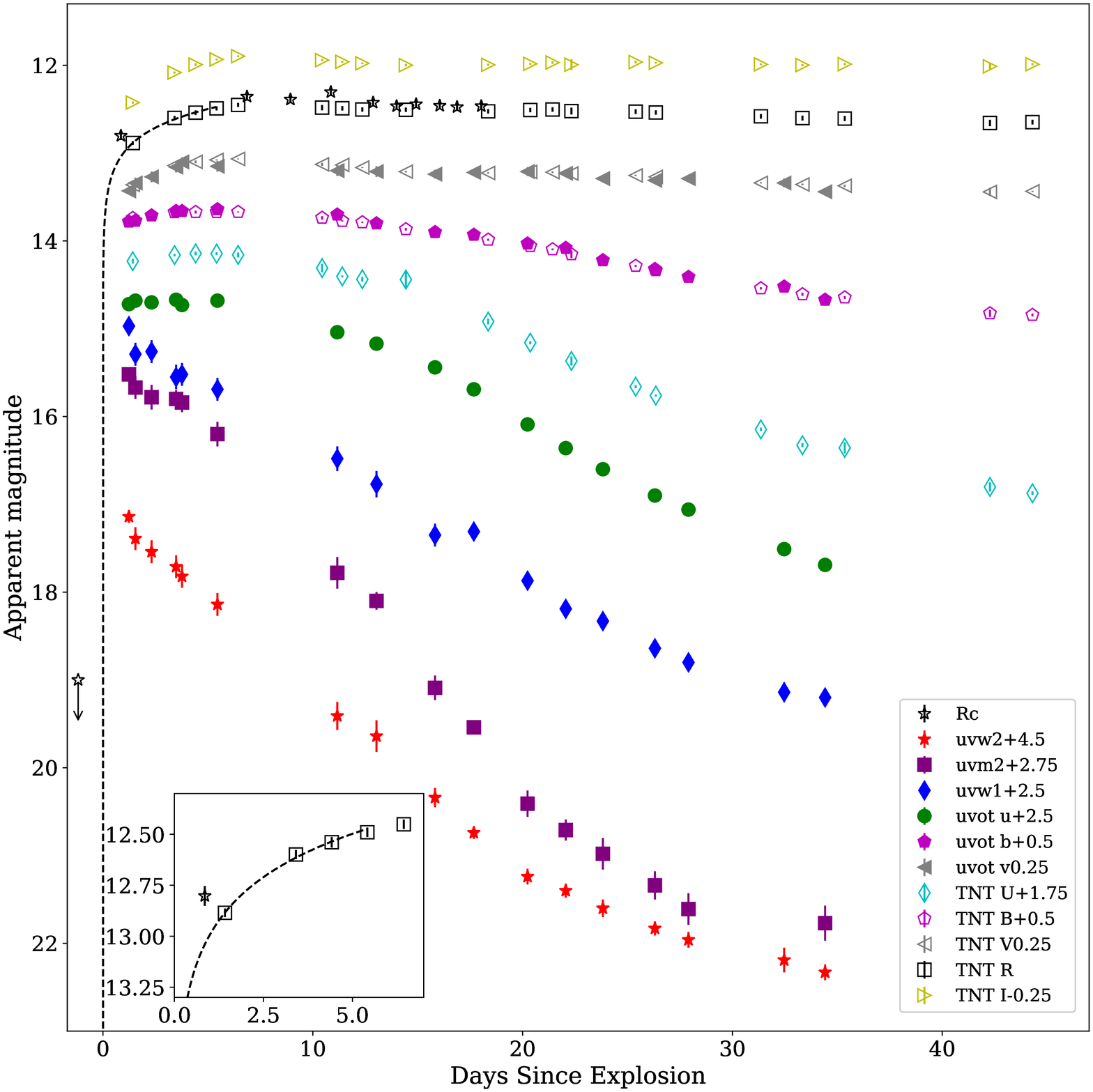}
	\caption{Optical and Ultraviolet follow-up photometry of SN 2017eaw obtained with the Tsinghua-NAOC 0.8-m telescope (TNT) and the Swift UVOT. The dark dashed lines show the best fit to the R-band data in a form of $f\propto (t-t_0)^n$, and the first open star represents the last non-detection(unfiltered) from Patrick Wiggins.}
	\label{fig:photometry}
\end{figure*}

\begin{figure*}
	\centering
	\includegraphics[width=0.9\linewidth]{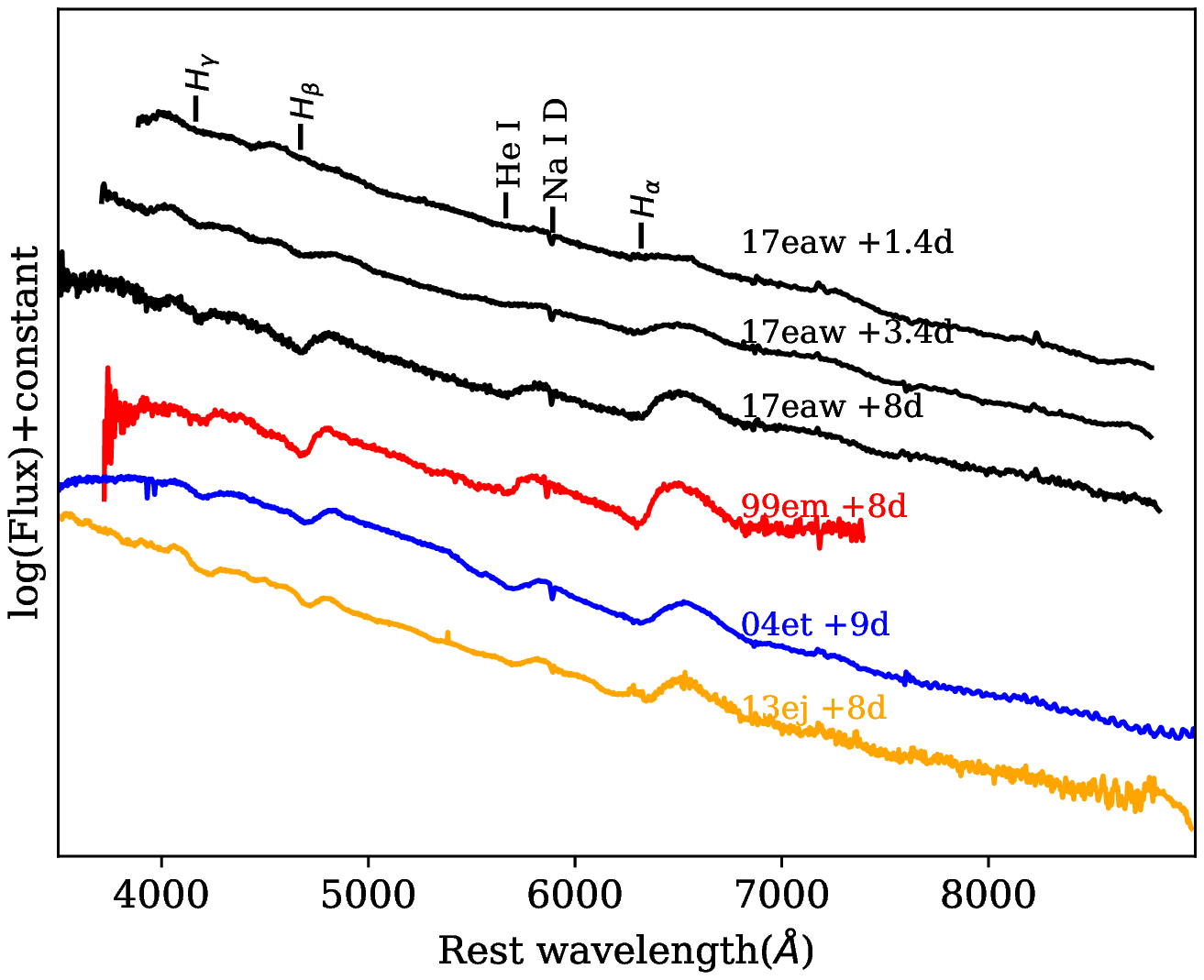}
	\caption{The spectra of SN 2017eaw in the early phase, and comparison with those of other well-studied SNe IIP around the maximum.}
	\label{fig:spec-sn}
\end{figure*}

Figure \ref{fig:photometry} displays the evolution of the UV and optical photometry of SN 2017eaw up to $\sim$45 days after the discovery. The unfiltered light curve derived from the observations by Patrick Wiggins is overplotted. The first nondetection can be traced back to 2017 May 12.20 UT, with an unfiltered CCD magnitude upper limit of about 19.0 mag. The explosion time (or first light) can be estimated as JD 2457886.72$\pm$1.01 by taking the mean point between the first detection and the last non-detection. The rising phase was well captured in the $UBVRI$ bands, while the UV data already started to decline from the first data points. We fit the TNT $R$-band light curve during the first week after explosion with a $f \propto (t-t_0)^n $ function, and find that the first (unfiltered) discovery point is brighter than the fitted curve by about 0.2 mag. After normalizing the unfiltered magnitudes to the R-band values in the plateau phase, this difference is about 0.16 mag, which could be related to the emission of shock breakout cooling tail. Such post-shock cooling decline before reaching the optical peak has been reported for SN 2014cx \citep{2016ApJ...832..139H}, SN 2016X \citep{2018MNRAS.475.3959H}, SN 2006bp \citep{2007ApJ...666.1093Q}, KSN 2011d \citep{2016ApJ...820...23G}. We caution, however, that the evidence for shock cooling detection in SN 2017eaw should be less significant due to that the difference between the unfiltered and R-band magnitudes change with time as a result of evolution of the spectral energy distribution of SNe IIP in the early phase.

By fitting a low-order polynomial to the data around maximum light, the supernova reached the $U$- and $B$-band peak of 12.39 mag and 13.16 mag at t$\sim$+5.0 days and t$\sim$+5.1 days from the discovery respectively. We note that the rise time in $B$ band is almost the shortest among a large sample of SNe II \citep{2015MNRAS.451.2212G}.

For the spectrum of SN 2017eaw displayed in Fig. \ref{fig:spec-sn}, the first two were obtained with BFOSC and the third one was obtained with OMR on the Xinglong 2.16-m telescope. All spectra were reduced using standard IRAF routines. Flux of these spectra were calibrated using nearby spectrophotometric standard stars with similar airmass. Atmospheric extinction was corrected basing on the extinction curve of Xinglong observatory.

The earliest spectrum, obtained at $\sim$1.4d after explosion, shows a blue continuum and weak features of Balmer lines that are characteristic of young type II supernovae \citep{2017CBET.4391....4X}. These Balmer lines become dominant features in the following spectra, similar to those seen in SNe IIP(Fig. \ref{fig:spec-sn}). Broad absorption feature of He I $\lambda$ 5876 become visible in the $+$3.4d spectra. The spectrum of SN 2017eaw at t$=+$8d is similar to other well-studied type IIP supernova(especially SN 2004et), while showing weaker and broader profiles of Balmer and He I lines compared to SN 1999em. Noticeable narrow interstellar Na I D absorption features can be seen at around 5892.5\AA\ in the spectra, with an equivalent width (EW) of 1.6$\pm$0.4\AA. Owing to the very close distance of the SN and the low resolution of our spectra, it is difficult to separate the absorption component due to the Milky way from that due to NGC 6946. The reddening towards SN 2017eaw can be estimated with the strength of this interstellar absorption line. Based on the existing empirical relation between the EW of Na ID absorption and dust extinction \citep{1990AA...237...79B, 2014MNRAS.442..844F}, the total extinction of A$_V$=1.24$\pm$0.30 or 2.41$\pm$0.62 mag can be inferred for SN 2017eaw. Due to this large discrepancy, we adopted an average value of A$_V$=1.83$\pm$0.59 mag for the extinction correction of the progenitor. Given the Galactic extinction of A$_V$ =0.94 \citep{2011ApJ...737..103S} towards SN 2017eaw, the host-galaxy extinction is estimated as A$_V$=0.89$\pm$0.59 mag and it is adopted in the progenitor analysis.

\section{Identification of the Progenitor Candidate}
\label{sec:proobs}
\subsection{Images from Hubble Space Telescope}
\label{sec:hst}

\begin{table*}
	\centering
	\caption{Log of the pre- and post-explosion HST images covering the explosion site of SN 2017eaw}\label{tab:hst-obs}
	\begin{tabular}{ccccccl}
		\hline
		Instrument & Filter & Obs date & Exposure time & Prospsal ID & PI & Drizzled image$^{a}$\\
		\hline
		ACS/WFC   & F814W & 2016-10-26 & $600 + 590 + 690 \times 2$ & 14786 & Williams & jd9g03020\_drc.fits \\
		ACS/WFC   & F606W & 2016-10-26 & $530 + 520 + 690 \times 2$ & 14786 & Williams & jd9g03010\_drc.fits \\
		ACS/WFC   & F814W & 2004-07-29 & 120             & 9788  & Ho      & hst\_9788\_d8\_acs\_wfc\_f814w\_drz.fits \\
		ACS/WFC   & F658N & 2004-07-29 & $350 \times 2$  & 9788  & Ho      & j8mxd8010\_drz.fits \\
		WFC3/IR   & F164N & 2016-10-24 & $599 \times 4$  & 14638 & Long    & id6o03010\_drz.fits \\
		WFC3/IR   & F160W & 2016-10-24 & $99  \times 4$  & 14638 & Long    & id6o03020\_drz.fits \\
		WFC3/IR   & F128N & 2016-02-09 & $353 \times 4$  & 14156 & Leroy   & icyqe2030\_drz.fits \\
		WFC3/IR   & F110W & 2016-02-09 & $228 \times 2$  & 14156 & Leroy   & icyqf2010\_drz.fits \& icyqe2040\_drz.fits \\
		WFC3/UVIS & F814W & 2017-05-29 & $27  \times 10$ & 14645 & van Dyk & id9702010\_drz.fits \\
		\hline
	\end{tabular}
	\begin{flushleft}
		$^{a}$ The retrieved drizzled images used in geometric transformation and finding of the progenitor of SN 2017eaw.
	\end{flushleft}
\end{table*}

Pre-explosion images of SN 2017eaw are available from Mikulski Archive for Space Telescopes (MAST) and the Hubble Legacy Archive (HLA). By searching the publicly available archive images, we found there are F128N- and F110W-band (WFC3/IR) images that were taken on 9 February 2016 (PI Leroy), F164N- and F160W-band images taken on 24 October 2016 (PI Long), and F606W- and F814W-band (ACS/WFC) images taken on 26 October 2016 (PI Williams). Besides many images taken in 2016, there are also F658N- and F814W-band (ACS/WFC) images taken on 29 July 2004 (PI Ho). Table \ref{tab:hst-obs} list the observation log of these images.
\begin{figure*}
	\centering
	\includegraphics[width=0.7\linewidth]{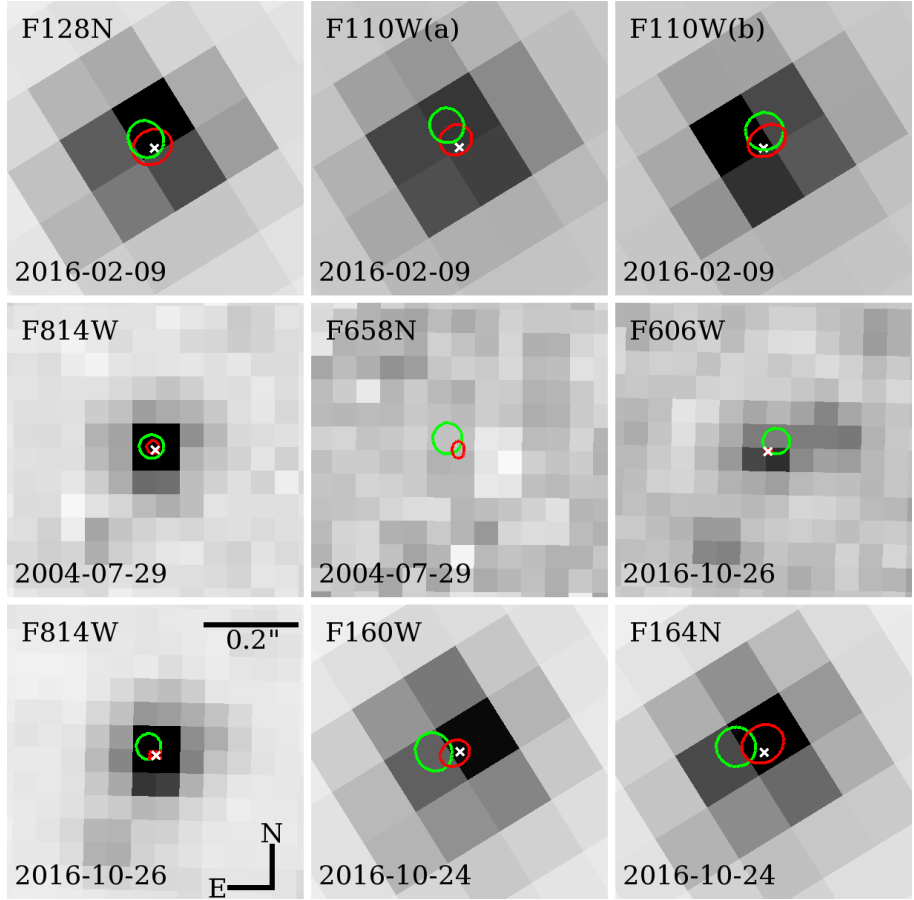}
	\caption{The zoomed-in region of the progenitor site on the HST pre-explosion images. All images are aligned. The white cross marks the measured center of the possible SN progenitor. The center of the ellipses shows the transformed SN positions, with the size of indicating the uncertainty. Red ellipses show the SN position determined from the post-explosion HST image, while the green ones represent the ones from the post-explosion TNT image.
	F110W(a): F110W image of icyqf2010\_drz.fits;
	F110W(b): F110W image of icyqe2040\_drz.fits.}
\label{fig:hst}
\end{figure*}

\begin{table*}
	\centering
	\caption{Transformed SN positions on pre-explosion HST images using post-explosion TNT image}\label{tab:position}
	\begin{tabular}{cccccc}
		\hline
		image & F128N &  F110W(a)$^{*}$ &  F110W(b)$^{**}$ &  F814W(2004) &  F658N\\
		\hline
		$s_\mathrm{pre}/(\mathrm{arcsec \cdot pixel^{-1}})^{a}$ & 0.13 & 0.13 & 0.13 & 0.05 & 0.05 \\
		$N_\mathrm{stars}^{b}$ & 25&24&26&9&14 \\
		$s_\mathrm{post}/(\mathrm{arcsec \cdot pixel^{-1}})^{c}$ & 0.52 & 0.52 & 0.52 & 0.52 & 0.52 \\
		$\sigma_\mathrm{total}/\mathrm{mas}^{d}$ & 37/41&36/38&41/40&26/25&31/33 \\
		$(x, y)_\mathrm{SN}^{e}$ & (558.28, 650.24)&(569.27, 656.08)&(539.49, 633.24)&(3360.32, 3978.51)&(877.62, 267.01) \\
		$(x, y)^{*f}$ & (558.23, 650.44)&(569.28, 656.50)&(539.65, 633.47)&(3360.48, 3978.36) & $\cdots$ \\
		\hline
	\end{tabular}
	\begin{tabular}{cccccc}
\hline
image                                                & F160W            &  F164N          &  F606W           &  F814W(2016) \\
\hline
$s_\mathrm{pre}/(\mathrm{arcsec \cdot pixel^{-1}})^{a}$  & 0.13             & 0.13            & 0.05             & 0.05             \\
$N_\mathrm{stars}^{b}$                                   & 18               & 18              & 15               & 16               \\
$s_\mathrm{post}/(\mathrm{arcsec \cdot pixel^{-1}})^{c}$ & 0.52             & 0.52            & 0.52             & 0.52             \\
$\sigma_\mathrm{total}/\mathrm{mas}^{d}$                 & 42/38            & 42/42           & 27/29            & 27/27            \\
$(x, y)_\mathrm{SN}^{e}$                                 & (935.82, 600.28) &(935.91, 600.43) &(2809.77, 989.06) &(2809.59, 990.81)\\
$(x, y)^{*f}$                                         & (936.04, 600.64) &(936.22, 600.78) &(2809.33, 989.40) &(2809.23, 990.51)\\
\hline
	\end{tabular}
	\begin{flushleft}
		$^{*}${F110W image of icyqf2010\_drz.fits.}\\
		$^{**}${F110W image of icyqe2040\_drz.fits.}\\
		$^{a}${Pixel scale of the pre-explosion image.}\\
		$^{b}${Number of stars used in the geometric transformation.}\\
		$^{c}${Pixel scale of the post-explosion image.}\\
		$^{d}${Total uncertainty of the transformed SN position on the pre-explosion images, with the values separated by the slash representing uncertainties along X-axis and Y-axis, respectively.}\\
		$^{e}${The transformed SN position of pre-explosion image.}\\
		$^{f}${The position of center of the nearby star.}\\
	\end{flushleft}
\end{table*}

To measure the SN position on these HST images, we used $I$-band image of SN 2017eaw taken by the Tsinghua-NAOC 0.8-m telescope (TNT) on 17 May, which has a typical full-width at half maximum (FWHM) of $1.6''$. The position of SN 2017eaw was determined by averaging the results from six different methods (centroid, Gaussian and ofilter center algorithm of IRAF task phot, IRAF task imexamine, daofind, and SExtractor) as $(735.154\pm0.007, 644.454\pm0.011)$. The uncertainty of this position is estimated as the standard deviation of the mean value. Several stars commonly seen on the HST images and the $I$-band TNT image are used to establish the transformation function (a 2nd-order polynomial) converting the coordinates of the $I$-band TNT image to that of the HST images. The resulting positions and uncertainties are listed in Table \ref{tab:position}. The uncertainties include errors in SN position and geometric transformation. The reference stars used in position transformation varies in different HST wavebands, and the exact number is listed as $N_\mathrm{stars}$ in Table \ref{tab:position}.

Figure \ref{fig:hst} shows the region centering on the SN site with the transformed position marked. A point-like source can be detected near the position of the SN on the F128N-, F110W- and F814W-band images, while no detection on the F658N-band image. Difference between the positions of SN and this point-like source is generally less than the uncertainty of the transformed positions.

On 29 May 2017, an F814W-band image of the SN was taken again by the HST with WFC3/UVIS (PI Van Dyk), which became publicly available immediately. We retrieved the image and measured the SN position, with the X-pixel being at 547.90$\pm$0.02 and Y-pixel being at 576.79$\pm$0.04. This pixel position is then transformed at the SN positions on the pre-explosion HST images (see also Figure \ref{fig:hst}), which has higher precision but is overall consistent with those derived from the TNT images. Based on these two results, we concluded that the point-like source marked in Figure \ref{fig:hst} is the candidate of the progenitor of SN 2017eaw.

\begin{table*}
	\centering
	\caption{Photometry of the Progenitor Star Identified for SN 2017eaw.}\label{tab:pro-log}
	\begin{tabular}{ccccc}
		\hline
		Telescope & Instrument & Filter &Exposure time(s)& m$^{a}$ \\
		\hline
		HST     & ACS/WFC     & F606W                 &530+520+690 $\times$ 2
		& 26.419(053) \\
		HST     & ACS/WFC     & F658N                 &350$\times$2
		& $>$23.7      \\
		HST     & ACS/WFC     & F814W(2004)           &120					  & 22.558(035) \\
		HST     & ACS/WFC     & F814W(2016)           &530+520+690 $\times$ 2 & 22.845(008) \\
		HST     & WFC3/IR     & F110W                 &228 $\times$ 2         & 20.344(005)  \\
		HST     & WFC3/IR     & F128N                 &353 $\times$ 4         & 19.749(019) \\
		HST     & WFC3/IR     & F160W                 &99 $\times$  4         & 19.351(007)  \\
		HST     & WFC3/IR     & F164N                 &599 $\times$ 4         & 19.023(012) \\
		Spitzer & IRAC        & $3.6\ \mathrm{\mu m}$ &2144                   & 18.02(070)  \\
		Spitzer & IRAC        & $4.5\ \mathrm{\mu m}$ &2144                   & 17.75(100)  \\
		Spitzer & IRAC        & $5.8\ \mathrm{\mu m}$ &2144                   & $>$17.6 \\
		Spitzer & IRAC        & $8.0\ \mathrm{\mu m}$ &2144                   & $>$16.9   \\
		Spitzer & MIPS        & $24\ \mathrm{\mu m}$  &2144                   & $>$12.6   \\
		\hline
	\end{tabular}
	\begin{flushleft}
		$^{a}$ The values listed in the parentheses represent the 1-$\sigma$ uncertainty of magnitude, in unit of 0.001 mag. The upper limit listed for some observations represents the 5-$\sigma$ detection limit.
	\end{flushleft}
\end{table*}

Photometry of the pre-explosion HST images is conducted with the DOLPHOT 2.0\footnote {\url{http://americano.dolphinsim.com/dolphot/}} package on the bias-subtracted, flat-corrected, so-called FLT FITS images that are retrieved from the MAST. Photometry is simultaneously measured using all ACS/WFC images and WFC3/IR images, following the routines that include the selection of a deep drizzled image as reference image, masking of bad pixels, calculations of sky background, alignment of images to the reference, and adoption of the recommended photometry parameters as described in the \emph{User's Guide} of DOLPHOT. Magnitudes and their uncertainties of the progenitor candidate are extracted from the output of DOLPHOT. The Magnitudes are PSF magnitudes, and all in Vega system. An upper limit of magnitude is given for the non-detection F658N image. The measured magnitudes of the SN progenitor in different wavebands and phases are reported in Table \ref{tab:pro-log}, with F606W-band and F814W-band magnitudes being 26.419 mag and 22.845 mag. The magnitudes of the SN progenitor were also independently measured by \cite{2018MNRAS.481.2536K} using the published HST images. In F606W, F814W, F160W bands, our results are consistent with theirs within 0.05 mag, while in F110W and F164N their results are fainter than ours by 0.4 mag and 0.1 mag, respectively.

\subsection{Images from the Spitzer Telescope}
\label{sec:spitzer}

\begin{figure*}
	\centering
	\includegraphics[width=0.7\linewidth]{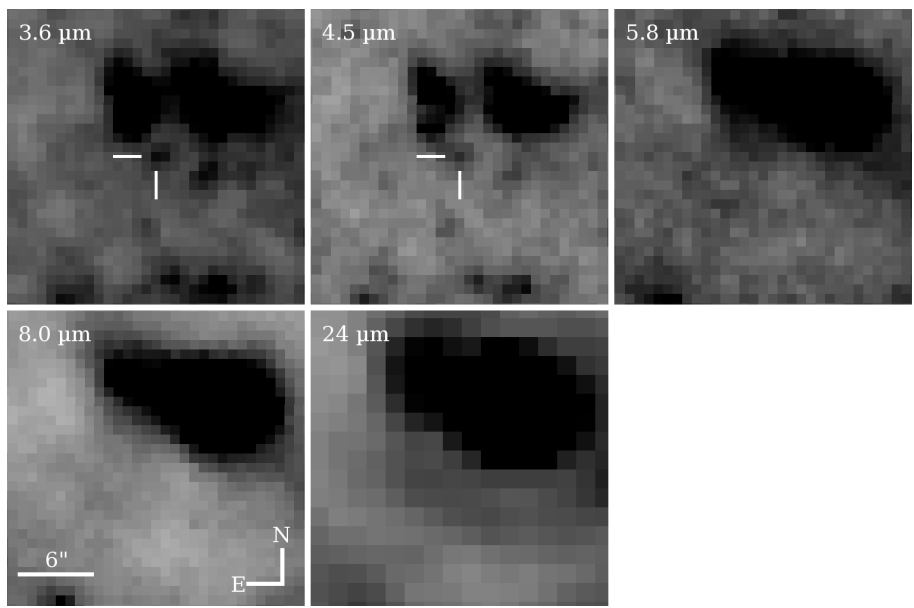}
	\caption{Subsections of pre-explosion Spitzer images centering on the site of the SN. All images are aligned. The white cross hairs in the $3.6\ \mathrm{\mu m}$ and $4.5\ \mathrm{\mu m}$ images mask the progenitor candidate.}
	\label{fig:spitzer}
\end{figure*}

The position calibration step was skipped for the Spitzer images, as the retrieved WCS information is well consistent with that from the GSC-2.3 standard catalogue. We find the counterpart of the progenitor candidate at the expected coordinates on $3.6\ \mathrm{\mu m}$ and $4.5\ \mathrm{\mu m}$ images (Figure \ref{fig:spitzer}). On the $5.8\ \mathrm{\mu m}$, $8.0\ \mathrm{\mu m}$ and $24\ \mathrm{\mu m}$ images, there are no clear point-like sources at the SN site. Photometry of the progenitor on Spitzer images is directly taken from \citep{2017ApJS..228....5K}, and  the limiting magnitudes of non-detection images are estimated using the same method. It should be pointed out the flux at the SN site of these images could be seriously contaminated by nearby sources that are not related to the progenitor of the SN\citep{2017ATel10373....1K} due to the low angular resolution of the Spitzer images. For example, the photometric aperture is adopted as 2.4" for the 3.6um/4.5um images\citep{2017ApJS..228....5K}, which corresponds to 48 pixels on the HST ACS images and 19 pixels on the WFC3/IR images, respectively. Adopting this large aperture for the HST images would result in an overestimate of the F814W-band magnitude by about 1.0 mag (i.e., $\sim$21.8 mag). Thus, we caution the use of the Spitzer magnitudes to constrain the progenitor properties of SN 2017eaw.

\section{Analysis}
\label{sec:alys}
\subsection{Constraining Properties of the Progenitor}
\label{sec:pro}

\begin{figure}
	\centering
	\includegraphics[width=\linewidth]{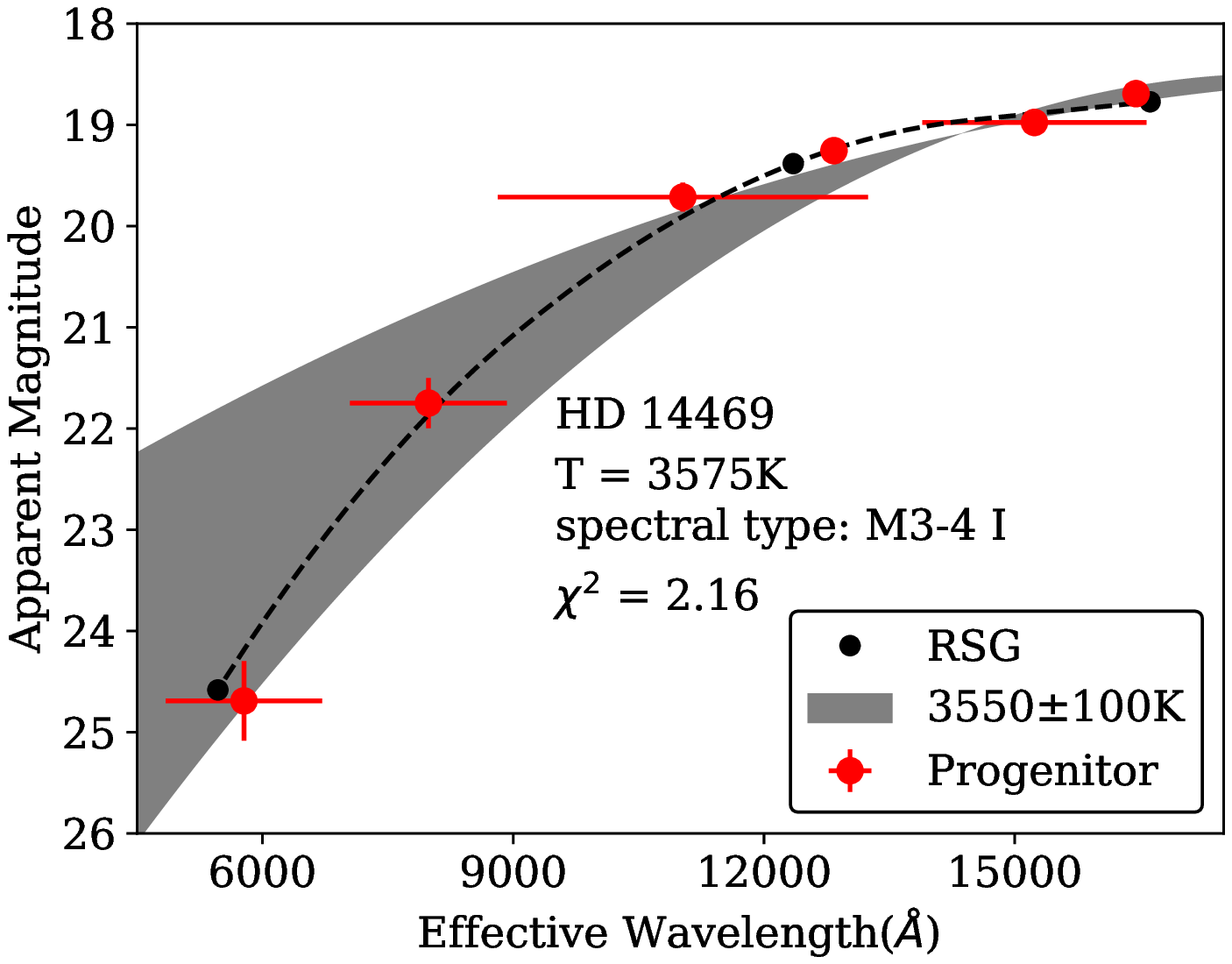}
	\includegraphics[width=\linewidth]{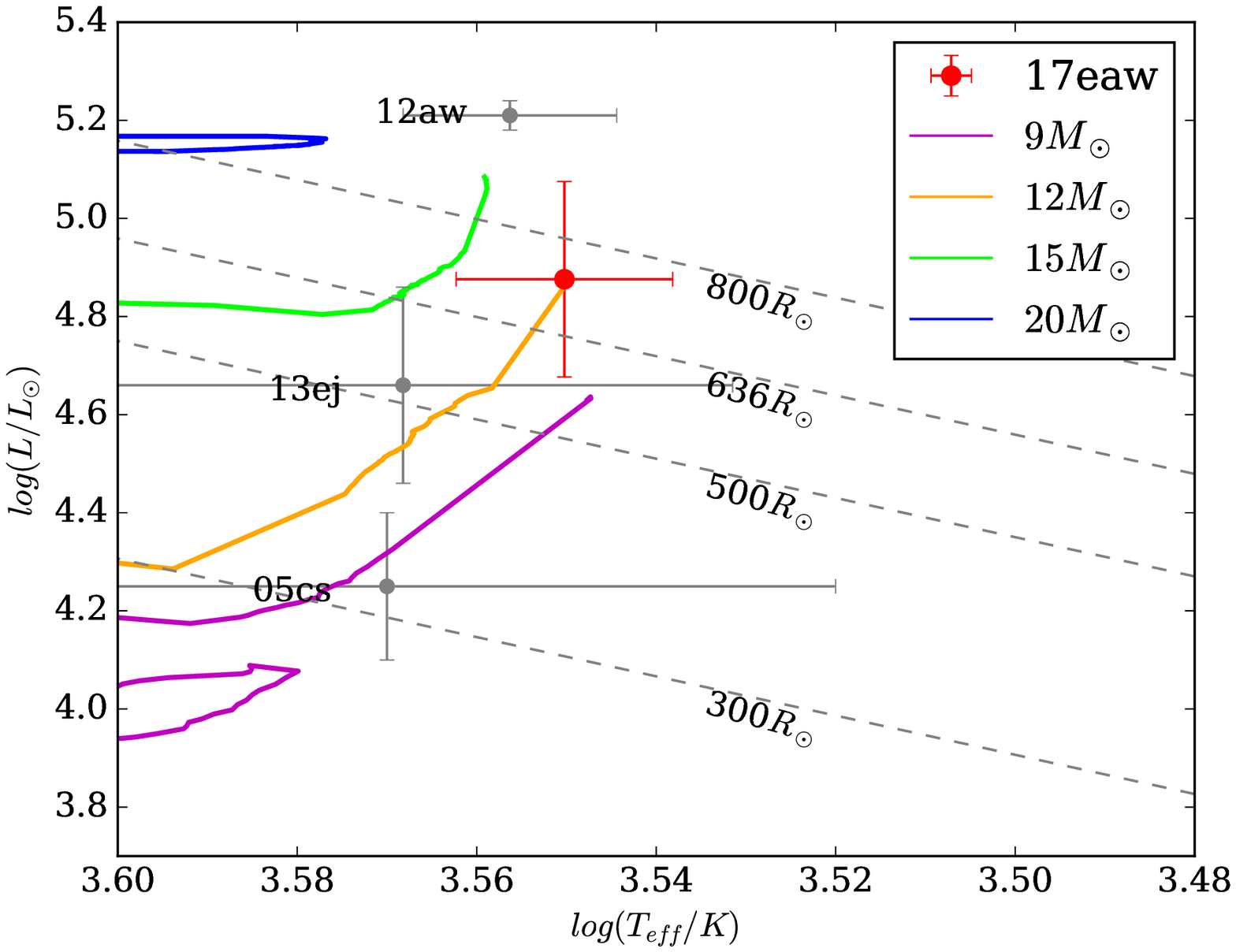}
	\caption{Spectral energy distribution (SED) of the progenitor star and its location in Hertzsprung-Russell diagram. Upper panel: the SED of the progenitor star compared to that of the Galactic Red Supergiants (RSGs). The red points represent the SED of the progenitor of SN 2017eaw obtained in 2016, while the black dots represent the ones of red supergiants from Milky Way. The black dashed line is yielded by applying a second-order polynomial fit to the values of Galactic RSGs. The gray shadow region represents the distribution inferred from the Galactic RSGs with the temperature ranging from 3450 K to 3650 K. Lower panel: the solid lines represent the stellar evolution tracks for 9$-$20 M$_\odot$, based on the Geneva model \citep{2012AA...537A.146E}. The measurements of the progenitors of SN 2005cs \citep{2005MNRAS.364L..33M} and SN 2013ej \citep{2014MNRAS.439L..56F} are over-plotted for comparison. The dashed straight lines indicate the lines of equal stellar radii.}
	\label{fig:track}
\end{figure}

SN 2017eaw provides a good opportunity to study the progenitor of SNe IIP, as the pre-explosion Hubble Space Telescope (HST) images are available in both optical and near-infrared bands. Inspecting the HST WFC3 images taken in 2016 reveals that a point source can be identified at the SN position in the F606W-, F814W-, F110W-, F128N-, F160W-, and F164N-bands (see Fig. \ref{fig:hst} and Sec. \ref{sec:hst} for details of the progenitor identification). This point source tends to become progressively brighter at longer wavelengths, consistent with a very red star (see Table 1 and Fig.\ref{fig:hst}). In the F814W band, the SN progenitor is measured to have a magnitude of 22.845$\pm$0.008 mag, which corresponds to an absolute magnitude of $-$6.93$\pm$0.51 mag after adopting a distance of 5.5 Mpc ($\mu$=28.67$\pm$0.43) and A$_{V}$ = 1.83$\pm$0.59 mag. This brightness is comparable to that of a red supergiant star. Such identification is also favored by the detection of a luminous and red source in the archival \textit{Spitzer} images at $3.6\ \mathrm{\mu m}$ and $4.5\ \mathrm{\mu m}$ wavelengths (see Fig.\ref{fig:spitzer}). The point source has an absolute magnitude of $-$10.65$\pm$0.38 mag at IRAC [3.6$\mu$ m] and a color of 0.27$\pm$0.06 in IRAC [3.6$\mu$ m]-[4.5$\mu$ m], which are consistent with the typical values of RSGs\citep{2010AJ....140...14S}. Note, however, that the nearby red stars could contaminate the Spitzer magnitudes due to the low angular resolution of the Spitzer images.

It is interesting that the region including the site of SN 2017eaw was also observed in 2004 by the HST Advanced Camera for Surveys (ACS). These observations were made in the F658N and F814W bands, where the progenitor can be clearly detected in F814W with m$_{F814W}$ = 22.558$\pm$0.005 mag. In comparison with this value, the SN progenitor appears dimmer by $\sim$0.29 mag in 2016, which is likely related to circumstellar dust formed around the progenitor shortly before its explosion (see discussions below). Note that the non-detection in the F658N band is due to the relatively short exposure time and narrow bandwidth of the filter. The measured magnitudes of the SN progenitor in different wavebands and phases are reported in Table \ref{tab:pro-log}.

To determine the spectral type of the progenitor of SN 2017eaw, we use 85 observed RSGs with effective temperatures ranging from 3325$K$ to 4200$K$ to fit the observed spectral energy distribution (SED). This sample consists of 74 Galactic RSGs\citep{2005ApJ...628..973L}, 7 RSGs in the Large Magellanic Cloud (LMC), and 4 RSGs in the Small Magellanic Cloud (SMC)\citep{2007ApJ...667..202L}.
A second-order polynomial is used to fit the $BVRIJH$-band magnitudes of the observed RSGs and then derive their SED curves. These curves are then compared with the reddening-corrected SED of the progenitor. The comparison shows that the progenitor matches well with an M4-type red supergiant with an effective temperature of about 3550$\pm$100 K (see the upper panel of Fig.\ref{fig:track}). In comparison, \cite{2018MNRAS.481.2536K} derived the effective temperature of 3350$^{+450}_{-250}$ K by fitting the SED of progenitor with stellar SED and dust models. Based on the derived absolute K-band magnitude (i.e., $-$10.33$\pm$0.20 mag) and the bolometric correction inferred from this band \citep{2006ApJ...645.1102L}, the bolometric magnitude or luminosity of the progenitor is derived as $-$7.45$\pm$0.50 mag or log($L_{bol}/L_\odot$) = 4.88$\pm$0.20. With the derived temperature and bolometric luminosity, the progenitor star can be well located in the Hertzsprung-Russel diagram. Inspection of stellar evolutionary tracks\citep{2012AA...537A.146E} shown in the lower panel of Fig.\ref{fig:track}, one can see that the progenitor of SN 2017eaw is likely to be a RSG star with an initial mass of 10 M$_\odot$ to 14 M$_\odot$ and a radius of about 700 R$_{\odot}$.

%Note that the temperature of the RSGs from literature is based on the TiO bands, which is usually lower than that inferred from the SED. Thus the radius derived here may be somewhat overestimated.

\begin{figure}
	\centering
	\includegraphics[width=\linewidth]{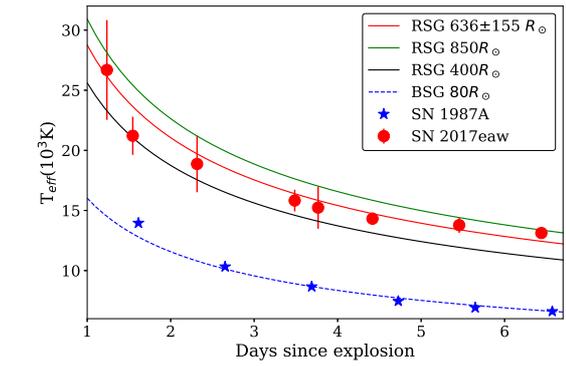}
	\caption{Constraining the progenitor radius of SN 2017eaw based on the cooling of the photospheric temperature in the early phase after the explosion. The red dots are the temperatures inferred from the blackbody fit to the observed UV and optical photometry; while the red curve represents the best-fit of the radius. The green and black curves show the upper and lower limits of the estimates of the radius, respectively. The blue stars show the temperature evolution of SN 1987A, and the blue dashed line shows the corresponding best-fit radius using the theoretical model \citep{2011ApJ...728...63R}.}
	\label{fig:rabinak}
\end{figure}

The radius of progenitor can be estimated from the evolution of photospheric temperature in the early stage of the explosion, which is primarily determined by the radius, opacity, and density profile of the progenitor \citep{2011ApJ...728...63R}. This early-time temperature evolution can be derived using the ultraviolet (UV) and optical photometry obtained within the first week after the explosion. Adopting the typical density profile as f$_\rho$ = 0.13 and Thomson scattering opacity as 0.34 cm$^2$ g$^{-1}$, the progenitor radius of SN 2017eaw can be determined as 636$\pm$155R$_{\odot}$ (Fig.\ref{fig:rabinak}). Moreover, the rise time of the light curve can be also used to estimate the radius of the progenitor of type IIP supernova through the relation between the progenitor radius and the rise time of the light curve, i.e., log R[R$_\odot$] = (1.225$\pm$0.178) log t$_{rise}$ [day] + (1.692$\pm$0.490) \citep{2016ApJ...829..109M}. For SN 2017eaw, the rise time is estimated to be t$_{rise}$ = 6.81$\pm$0.85 day in the g band, which gives an estimate of the radius as 515$\pm$180 R$_{\odot}$. The above two estimates of the progenitor radius agree within 1-$\sigma$ error.

\begin{figure}
	\centering
	\includegraphics[width=\linewidth]{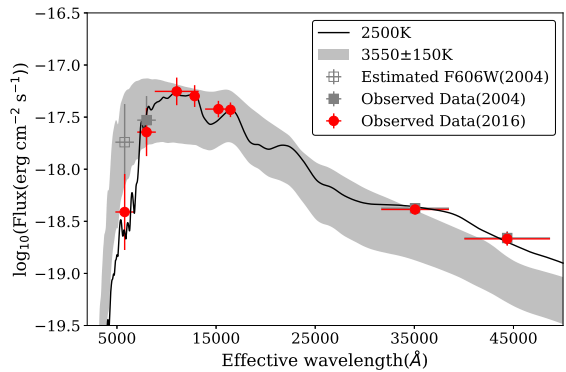}
	\caption{The gray squares and red circles are the spectral energy distribution of the progenitor observed in 2004 and 2016 respectively. The best fitting result of the observed SED in 2016 with theoretical MARCS stellar spectra indicates the progenitor to be a RSG with effective temperature of 2500K. The open square is the estimated F606W-band result in 2004. And the gray shadow region is the distribution of the SED inferred from the MARCS stellar spectra model with a radius of 636$\pm$155R$_\odot$.}
	\label{fig:marcs}
\end{figure}

We also try to fit the observed SED of the progenitor star from the prediscovery HST and \textit{Spitzer} images with synthesized spectra calculated using the stellar evolution code MARCS\footnote{\url{http:marcs.astro.uu.se/}}.
However, the best-fit model spectrum has an effective temperature of only about 2500 K (A similar value of 2600 K was also suggested by \cite{2018MNRAS.481.2536K}) with A$_V$=1.83$\pm$0.59 mag by including the Spitzer data(Fig.\ref{fig:marcs}), much lower than that inferred for the HST data alone. Given the luminosity, the lower temperature would result in an effective radius of 1330 R$_\odot$ for the progenitor, which is inconsistent with the estimate either from the rise time or temperature evolution of the shock cooling in the early phase. Despite an overall reasonable fit, the measurements in two Spitzer bands may suffer large contaminations of the nearby red sources due to large aperture photometry and/or the presence of circumstellar dust. In the latter case, the light of the progenitor star is scattered/absorbed by the surrounding dust and re-emitted at longer wavelengths (see also the analysis by \citep{2018MNRAS.481.2536K}). We also derive the F606W-band magnitude in 2004 by assuming that the dimming behavior in F814W-band is totally due to the CSM dust. Taking into account this speculated F606W-band magnitude and the observed F814W-band magnitude obtained in 2004, the SED of the progenitor inferred in 2004 matches with the theoretical stellar spectra model characterized by an effective temperature of 3550K$\pm$150K and a radius of 636$\pm$155R$_\odot$ (see Fig. \ref{fig:marcs}). As an alternative, the dimming behavior in F814W band could be also due to the intrinsic variation of a RSG star, but this explanation is inconsistent with either the observation taken at one year to a few days before the explosion or the theoretical prediction for its final stage evolution (see discussions in Sec \ref{sec:Var}).

\subsection{The Dust Around the Progenitor}
\label{sec:dust}

\begin{figure*}
	\centering
	\includegraphics[width=\linewidth]{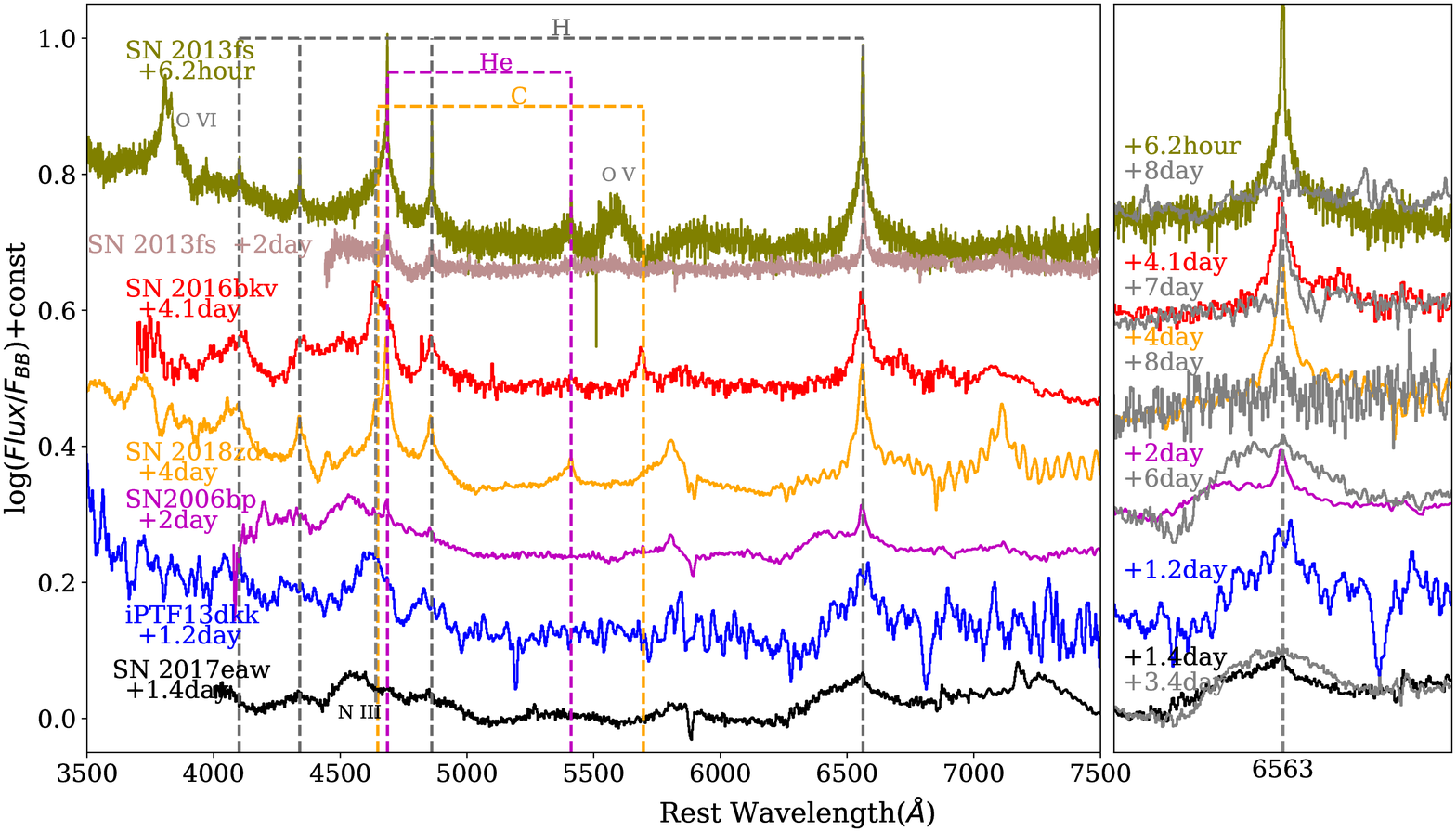}
	\caption{Early time spectroscopy of SN 2017eaw and some comparison SNe II that show clear signatures of circumstellar materials. Left panel: the early spectroscopy of SN 2017eaw, SN 2006bp \citep{2007ApJ...666.1093Q}, SN 2013fs \citep{2017NatPh..13..510Y}, iPTF13bkk \citep{2016ApJ...818....3K}, SN 2016bkv \citep{2018ApJ...859...78N}, and SN 2018zd (from our own database), covering the wavelength from 3500\AA\ to 7500\AA. All the spectra were obtained within 4 days after explosion and have been normalized to the continuums and corrected for the host-galaxy redshift. The positions of narrow emission lines of hydrogen, helium, and carbon are marked with dashed lines. Right panel: early-time spectral evolution centering at H$_\alpha$ line profile, with the ion-flashed/circumstellar features being initially detected and subsequently having disappeared. The red flux deficiency and blue excess asymmetry feature of H$_\alpha$ in the early spectra of SN 2017eaw also disappears in the later spectra (at t$\sim$+3.4day) making the line profile symmetric, which is a signature of dust.}
	\label{fig:spec}
\end{figure*}

\begin{figure*}
\centering
\includegraphics[width=\linewidth]{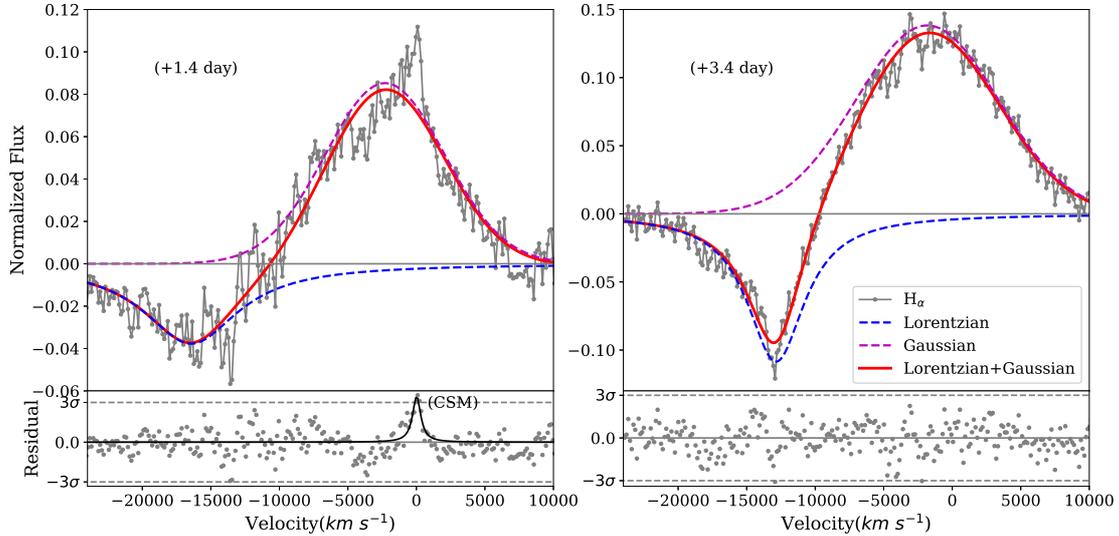}
\caption{The H$_\alpha$ line profile in the earliest two spectra of SN 2017eaw, displayed in a velocity space.Left panel: the H$_\alpha$ line profile in t=+1.4 day spectrum. The red curve represents the best-fit result using a combination of Lorentzian and Gaussian functions, with the blue and magenta curves showing the absorption and emission components of the P-Cygni profile, respectively. Right panel: the same case as in the left panel but for the t=+3.4 day spectrum. The lower panels show the residual of the observed line profile relative to the best-fit profile; and the narrow emission component in the left panel, due to the CSM, can be fitted by a Lorentzian function with a center wavelength of 6559.50$\pm$1.31\AA.}
\label{fig:Hdecomp}
\end{figure*}

\begin{figure}
\centering
\includegraphics[width=\linewidth]{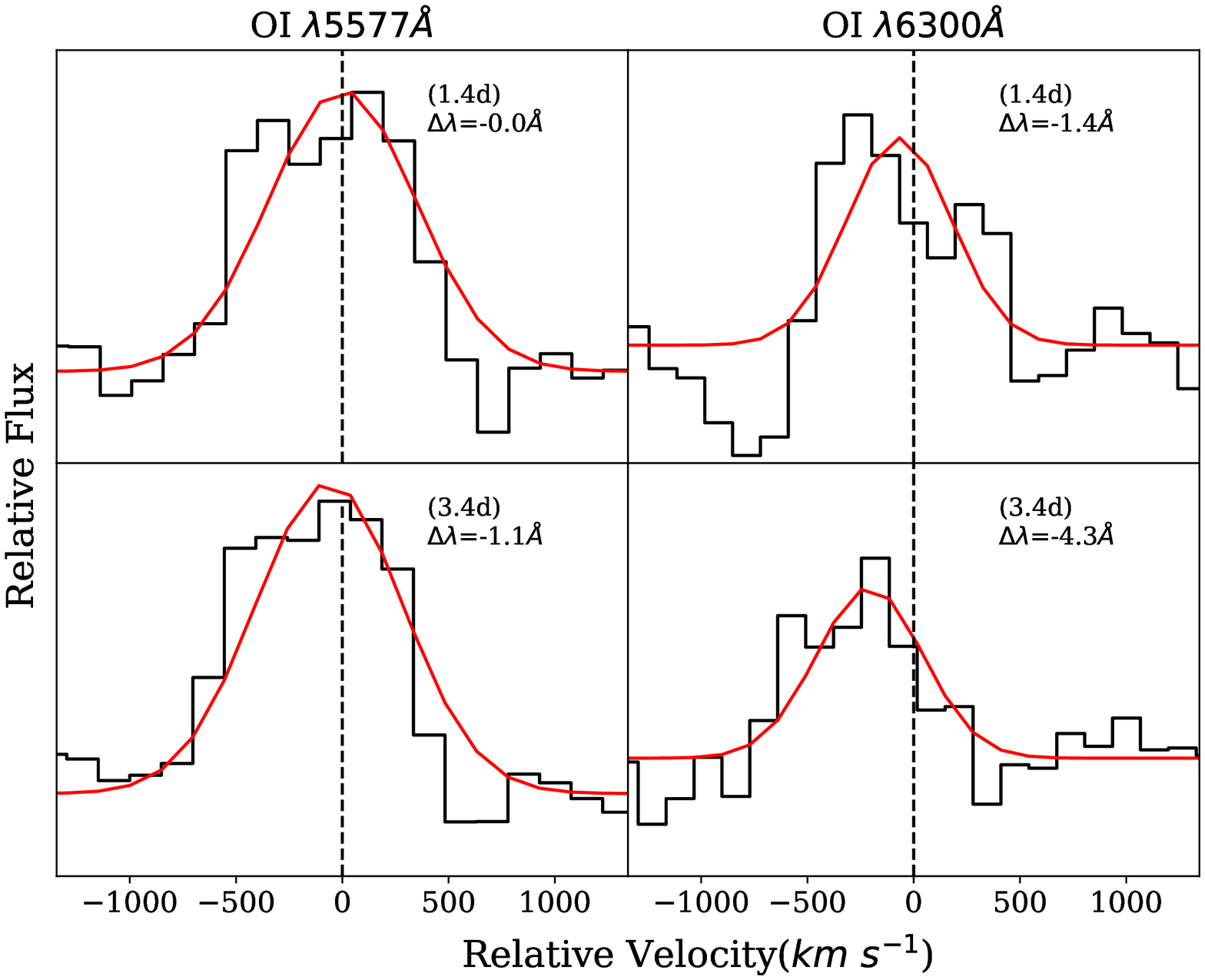}
\caption{[OI $\lambda\lambda$5577, 6300\AA] sky emission lines. The upper panel shows [OI $\lambda\lambda$5577, 6300\AA] sky emission lines for the t = $+1.4$ day spectrum, while the lower panel indicates the case for the t$\sim$3.4 days spectrum. The sky emission lines are not corrected for the redshift, and the red lines shows the result of Gaussian fit.}\label{fig:skylines}
\end{figure}

 Some recent studies reveal that the ion-flashed features can be detected in very early spectra of SN 2013fs \citep{2017NatPh..13..510Y}, and SN 2016bkv \citep{2018ApJ...859...78N}, which proves the existence of CSM around SNe IIP. The detection of prominent asymmetric H$_\alpha$ emission feature in the t=$+$1.4 day spectrum provides strong evidence for the existence of CSM dust around SN 2017eaw (see Fig.\ref{fig:spec}). This weak feature is also detected in a spectrum obtained an hour earlier \citep{2017ATel10374....1C}. The asymmetric H$_\alpha$ line profile seen in SN 2017eaw can be decomposed into two components, with the broad component formed in the SN ejecta and the weak narrow peak formed due to the ionization of the surrounding CSM. The weak emission feature is measured to have a wavelength of 6559.50$\pm$1.31\AA\ , corresponding to an expansion velocity of about 163 km s$^{-1}$ for the CSM, as shown in Fig.\ref{fig:Hdecomp}. This measurement of velocity is validated by carefully inspecting the position of sky emission lines [O I] $\lambda$5577 and [O I] $\lambda$6300 in the spectrum to ensure the accuracy of wavelength calibration (Fig.\ref{fig:skylines}). The stellar wind inferred from the ion-flashed signature in the spectrum of SN 2013fs shows a similar velocity (i.e., $\sim$100 km $^{-1}$) \citep{2017NatPh..13..510Y}.

Similar asymmetric H$_{\alpha}$ line profile can also be seen in the earliest spectra of SN 2006bp and iPTF13dkk \citep{2016ApJ...818....3K}, but only narrow emission features are shortly detected in the flash spectra of SN 2013fs, SN 2016bkv, and SN 2018zd. In the very early phase, the spectra of SN 2013fs and SN 2006bp also show prominent He II lines, while this feature is blended with C~III/N~III in iPTF13dkk, SN 2016bkv, and SN 2017eaw. Such an observed diversity of the flashed spectroscopy suggests that the CSM/dust environments around their progenitor stars are quite diverse.

The duration of narrow emission features also varies for different SNe II (see Fig.\ref{fig:spec}). For SN 2017eaw, the weak narrow H$_{\alpha}$ emission lines seen at t$\sim$+1.4 days disappeared in the t$\sim$+3.4 spectrum, suggesting that the obscuring dust shell should be thin and close to the progenitor star. Such a quick variation can be explained by that the CSM was initially ionized by the supernova photons, and then it was captured and destroyed by the fast-expanding SN ejecta. Assuming that the outermost layer of the SN ejecta has a velocity of $~$15,000-20,000 km s$^{-1}$ as inferred from the blue-shifted absorption of H$_{\alpha}$ and it swept the CSM within two days, and we estimate that the CSM has a distance $<$2.1-4.3$\times$10$^{14}$cm from the supernova. With this distance, one can speculate that the progenitor star ejected the CS materials at 1-2 years before the explosion with the derived stellar wind velocity of about 160 km s$^{-1}$.

\subsection{Variability of the progenitor}
\label{sec:Var}
\begin{figure*}
	\centering
	\includegraphics[width=\linewidth]{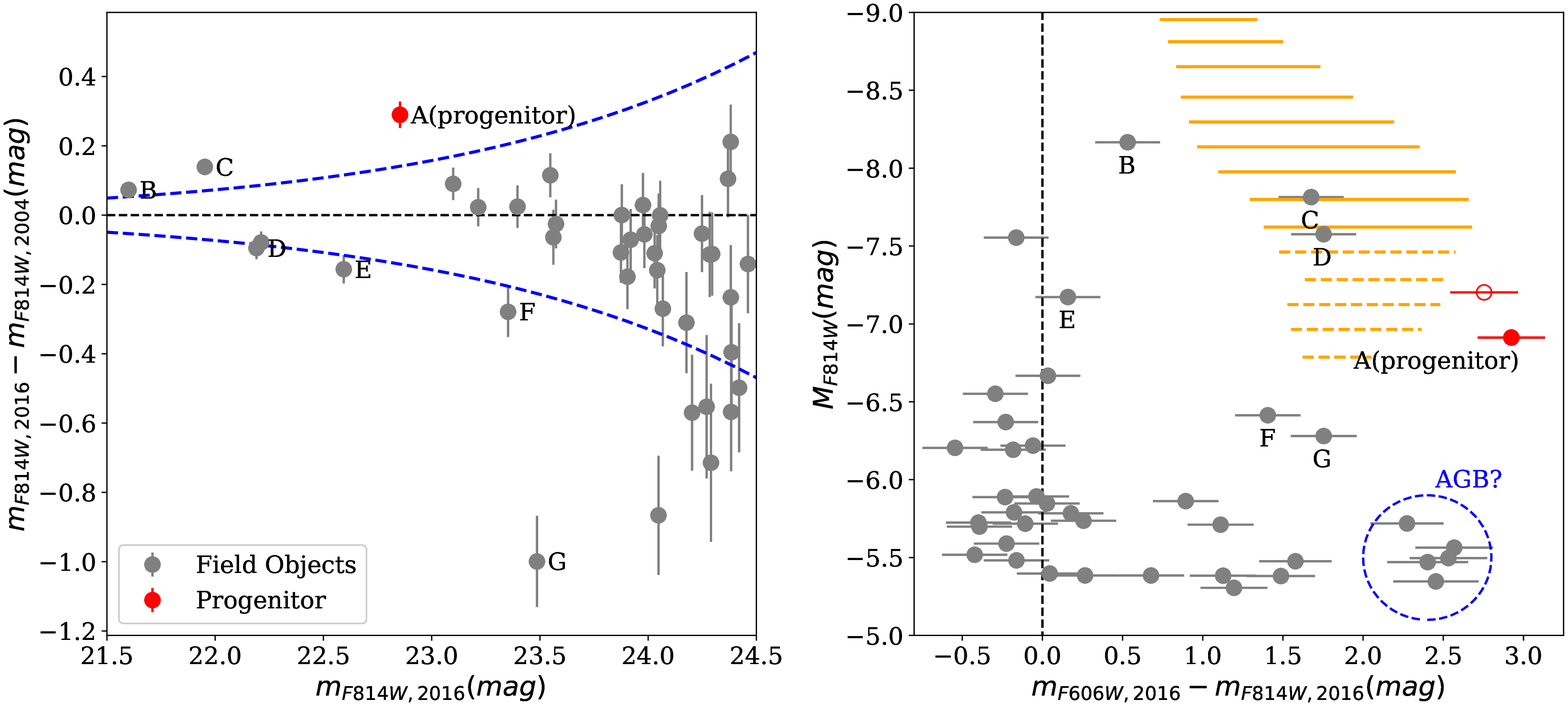}
	\caption{Magnitude and color variations for the progenitor star of SN 2017eaw and the neighboring stars at a distance of 200 pc, measured from the HST prediscovery images. Left panel: the F814W-band magnitude variation over the period from 2004 to 2016. The blue dashed lines indicate the 3-$\sigma$ limit of the photometric error. The progenitor of SN 2017eaw is represented by symbol "A", and those stars showing larger variations are also labeled with specific symbols. Right panel: the F606W-F814W color distribution of progenitor and field stars, measured from the HST images taken in 2016. The red open circle represents the color that the progenitor would have had in 2004 assuming that the observed dimming behavior is totally due to the obscuration by accumulated dust. The orange solid and dashed lines show the instability region of supergiants for stars in the mass range of 15 M$_\odot$ $<$ M $<$ 40 M$_\odot$ and 10 M$_\odot<$ M $<$15 M$_\odot$, respectively. These regions are obtained by applying the pulsation growth rate to the MIST stellar tracks\citep{2018arXiv180405860C}. The objects locating in the dashed circle are likely luminous asymptotic giant branch (AGB) stars according to their luminosity and red colors.}
	\label{fig:fieldobjects}
\end{figure*}

To understand the dimming behavior of the progenitor star in F814W-band over the period from 2004 to 2016, we examined in Figure \ref{fig:fieldobjects} the magnitude variation of the stars within 200 pc from the progenitor. On a timescale of a decade, most of the field stars are found to have light variations less than $~$0.3 mag when considering the larger photometric uncertainty at the faint end (i.e., m$_{F814W}>$24.0 mag; see left panel of Fig.\ref{fig:fieldobjects}); while the SN progenitor and star G are the two ones that show noticeable light variations with a relatively high significant level. Although a significant magnitude change might be common for a red supergiant star during its evolution, becoming faint is not predicted by theory for a red supergiant in its final stage towards explosion. In 2004 and 2016, the progenitor of SN 2017eaw should be near the end of carbon burning and in the neon/oxygen burning phase, respectively \citep{2005NatPh...1..147W}. In the neon/oxygen burning stage, the density wave energy is mainly deposited into the hydrogen envelope and consequently the surface luminosity will modestly increase instead of becoming faint \citep{doi:10.1093/mnras/stx1314}. Thus the dimming behavior observed for the progenitor of SN 2017eaw is unlikely due to the intrinsic variability of a RSG. Moreover, the Ks-band observations of the progenitor star with Keck/MOSFIRE and P200/WIRC, obtained at one year to a few days before the explosion, did not show significant light variation at a 6\% level \cite{2019arXiv190101940T}, which is consistent with our speculation that the light variation is not intrinsic to the red supergiant.

We further examined the distribution of the m$_{F606W}$-m$_{F814W}$ color as measured from the 2016 HST images. After corrections for the reddening, the progenitor is found to have a m$_{F606W}$-m$_{F814W}$ color of 2.94$\pm$0.20 mag, which is noticeably redder than the nearby RSGs. Its position is very close to the instability regime of supergiants with mass in a range of 10-15 M$_{\odot}$, in particular when considering the effect of CSM dust (see Fig.\ref{fig:fieldobjects}). On the other hand, instability mass ejection is also needed to reproduce the coolest (or reddest) variables with masses lower than 10-15 M$_{\odot}$ according to the census of luminous stellar variability in M51 \citep{2018arXiv180405860C}. In the supergiant instability regime, a pulsational superwind could lead to a dramatically enhanced mass loss at late phases, which may account for an ejection of the H-envelope of RSG stars just before their explosions \citep{2010ApJ...717L..62Y, 2018arXiv180405860C}.

\section{Conclusion}
\label{sec:sum}
From the above analysis, we conclude that the normal type IIP supernova 2017eaw has an M4-type red supergiant star with an initial mass 12$\pm$2 M$_{\odot}$, which further confirms the trend that most SNe IIP arise from core-collapse explosions of RSGs. However, the multi-epoch HST images obtained before the SN detection reveal that the RSG progenitor became faint by 30\% one year before it exploded. Although this dimming phenomenon could be due to the intrinsic light variation of the RSG star, detailed analysis of the neighbouring stars and early time spectra suggest that it is more likely due to the obscuration of circumstellar dust shell newly-formed at a distance of $<$2.1-4.3$\times$10$^{14}$ cm from the star. The fast-moving circumstellar materials (at a velocity of $\sim$ 160 km s$^{-1}$) indicate that they were ejected from the star in a violent manner within a few years prior to the explosion. Thus, the observations of SN 2017eaw indicates that some low-mass RSGs could also experience a strong mass loss, perhaps invoked by pulsational instability, during its final-stage evolution towards explosion.

\section*{Acknowledgements}
We thank Patrick Wiggins for useful unfiltered images. This work is supported by the National Natural Science Foundation of China (NSFC grants 11325313 and 11633002), and the National Program on Key Research and Development Project (grant no. 2016YFA0400803). J.-J. Zhang is supported by NSFC (grants 11403096, 11773067), the Youth Innovation Promotion Association of the CAS(grants 2018081), the Western Light Youth Project, and the Key Research Program of the CAS (Grant NO. KJZD-EW-M06). T.-M. Zhang is supported by the NSFC (grants 11203034). This work was also partially Supported by the Open Project Program of the Key Laboratory of Optical Astronomy, National Astronomical Observatories, Chinese Academy of Sciences. The research of JRM is supported through a Royal Society University Research Fellowship.

%%%%%%%%%%%%%%%%%%%%%%%%%%%%%%%%%%%%%%%%%%%%%%%%%%


\newpage
\bibliographystyle{mnras}
\bibliography{mnras_17eaw-0214} % if your bibtex file is called example.bib

% Don't change these lines
\bsp	% typesetting comment
\label{lastpage}
\end{document}